\documentclass[12pt]{article}
\pdfoutput=1
\usepackage{putex}
\usepackage{amsmath}
\usepackage{amssymb}
\usepackage{simpler-wick}
\usepackage{graphicx}
\usepackage{epstopdf}
\usepackage{enumerate}
\usepackage{cite}
\usepackage{tensor}
\usepackage{slashed}
\usepackage{feynmf}
\usepackage{hyperref}
\usepackage{bbold}
\usepackage{mathrsfs}
\usepackage{caption}
\usepackage{subcaption}
\usepackage{axodraw2}
\usepackage{pstricks}
\usepackage{color}
\usepackage{youngtab}
\usepackage{amsfonts}
\usepackage{braket}

\newcommand\numberthis{\addtocounter{equation}{1}\tag{\theequation}}

\usepackage[utf8x]{inputenc}
\usepackage{braket}
\usepackage{titlesec}
\usepackage{fixltx2e}
\usepackage{cleveref}
\hypersetup{breaklinks}
\usepackage{comment}

\newcommand {\be} {\begin {equation}}
\newcommand {\ee} {\end {equation}}
\newcommand{\bea}{\begin{eqnarray}}
\newcommand{\eea}{\end{eqnarray}}

\makeatletter
\renewcommand{\@maketitle}{
\newpage
 \begin{center}%
  {\large\bfseries \@title \par}%
 \end{center}%
 \par} \makeatother

\numberwithin{equation}{section}

\titleformat*{\section}{\large\bfseries}

\institution{KY}{Department of Physics and Astronomy, University of Kentucky, Lexington, KY, 40506}

\begin{document}

\title{Towards Color-Kinematics Duality in Generic Spacetimes}

	\authors{Allic Sivaramakrishnan\worksat{\KY} 
\let\thefootnote\relax\footnote{\texttt{${}$allicsiva@uky.edu}}
 }

\abstract{In this note, we study color-kinematics duality in generic spacetimes. We work with a contact representation for on shell correlators. The position-space integrand is encoded by enumerated differential operators. This setup generalizes certain features of S-matrix kinematics to curved space. Differences between flat and curved space are captured by commutators. We study the nonlinear sigma model at four points as an explicit example and find that color-kinematics duality holds in generic spacetimes. We illustrate our approach in the AdS transition amplitude, a type of on shell correlation function. We find a double copy procedure at four points that connects the nonlinear sigma model, the biadjoint scalar theory, and the special Galileon theory.
}

\date{}

\maketitle
\setcounter{tocdepth}{2}
\tableofcontents

\section{Introduction}

Color-kinematics duality and the double copy provide a remarkable bridge between gauge theory and gravity. For example, consider a tree-level scattering process in Yang-Mills theory. The amplitude can be written in terms of color factors $c_i$, kinematic numerators $n_i$, and the Mandelstams $s_{\alpha_i}$ that specify a cubic graph,
\begin{equation}
\mathcal{A}_{YM} = \sum_i \frac{c_i n_i}{\prod_{\alpha_i} s_{\alpha_i}}.
\end{equation}
The $n_i$ are functions of momenta and polarizations, but according to color-kinematics duality, they can be chosen to obey the Jacobi relations of their sibling color factors, $n_i=n_j+n_k$  and  $c_i = c_j+c_k$ respectively \cite{BCJ}. With numerators in this form, the double copy now emerges. The replacement $c_i \rightarrow n_i$ produces a four-point tree amplitude in gravity,
\begin{equation}
\mathcal{A}_{Gravity} = \sum_i \frac{n_i^2}{\prod_{\alpha_i} s_{\alpha_i}}.
\end{equation}
The first appearance of double copy structure was a relation between tree amplitudes of strings \cite{KLT}. Color-kinematics duality and the double copy were later discovered at high multiplicity and loop level in terms of kinematic numerators \cite{BCJ, BCJLoop}. Color-kinematics duality has since been found in a vast network of field theories that are related by the double copy, and a double copy has been found at the classical level as well \cite{BCJReview}. The double copy has been useful in practice, enabling recent progress on the UV properties of gravity \cite{FiveLoops,BernCCEJPRZ18} and on gravitational waves \cite{ BernCRSSZ19a,BernCRSSZ19b,BernPRRSSZ21}.

The origins of color-kinematics duality and the double copy have been studied in flat space scattering processes, both in field theory and string theory \cite{CHYa,CHYb,CasaliMT20, CKisResidueTheorem,Stieberger09,Stieberger21,MonteiroO11,ChenJTW19,ChenJTW21,CheungM21}. However, it remains unclear to what extent color-kinematics duality depends on Poincare invariance, or more generally, having a symmetric background. If color-kinematics duality and the double copy truly are intrinsic properties of certain quantum field theories, they must exist in some form in curved backgrounds that do not possess any spatial symmetry. To this end, it may be useful to evaluate the idea that
\begin{equation}
 \text{Color}+\text{On-shell external legs} \Rightarrow \text{Color-kinematics duality} 
\end{equation}
for certain theories in curved space that obey color-kinematics duality in the flat space limit. The above ingredients are by no means sufficient to explain color-kinematics duality even in flat space, although a recent approach centered on the equations of motion clarifies this issue \cite{CheungM21}. There can also be many ways to promote a flat space theory to curved space, so it is not always clear what curved-space theory to use. Although we will not resolve these issues in this note, we aim to initiate progress towards color-kinematics duality in general spacetimes by providing tools and basic results.

Progress on color-kinematics duality in non-trivial backgrounds has so far focused on certain flat-space examples and Anti-de Sitter space (AdS). Color-kinematics duality\footnote{Non-trivial color-kinematics duality first appears at four points, while the double copy can be studied at three points \cite{BzowskiMS17,FarrowLM18,LipsteinM19}.} has been explored recently in AdS \cite{EberhardtKM20,ArmstrongLM20, AlbayrakKM20,AldayBFZ21, Zhou21,DiwakarHRT21}, and related ideas have appeared in the context of plane wave backgrounds \cite{AdamoCMN17, AdamoCMN18,AdamoI20}, massive processes \cite{Momeni:2020vvr,MomeniRT20,JohnsonJP20}, and celestial amplitudes \cite{CasaliP20, PasterskiP20}. In more detail, the embedding space approach taken in \cite{EberhardtKM20, DiwakarHRT21} led to color-kinematics duality at six points, and the Mellin space approach in \cite{AldayBFZ21,Zhou21} led to a double copy for supersymmetric theories that relates four-point boundary correlators in AdS$_5 \times S^n$ at different $n$. The AdS momentum space approach in \cite{ArmstrongLM20, AlbayrakKM20} led to a double copy relation at four points between the biadjoint scalar and Yang-Mills theories that involved shifting the spacetime dimension. These approaches to double copy in AdS have made use of representations of the boundary correlation functions that are particularly simple for vacuum AdS/CFT. Approaches that leverage the special features of AdS/CFT may continue to be fruitful, and a similar approach may succeed in de Sitter \cite{GomezLJ21}.

In this note, we take a different but complementary approach. We investigate color-kinematics duality in spacetimes without any spatial symmetry. While our results hold in empty AdS, they are valid more generally. We now summarize our findings.

Not all spacetimes admit an S-matrix or a notion of non-interacting asymptotic states. In this work, we instead make progress by studying correlation functions in curved space. Time-ordered correlators in a weakly-coupled QFT are computed by Feynman diagrams.  At tree level, the external legs of these diagrams are Feynman propagators, $G_F(x_i,x)$, which are off shell. We can obtain a different object by replacing $G_F(x_i,x)$ with Wightman functions, $G^{\pm}(x_i,x)$, which are on shell. The resulting ``on shell correlator'' plays a role similar to that of the S-matrix.\footnote{We define this procedure only diagrammatically in weakly-coupled QFT, though a similar object can be defined non-perturbatively at least in certain cases \cite{ BalasubramanianGL99}.}. These on shell correlators first appeared in the context of general spacetimes in \cite{MeltzerS20}. In certain settings, on shell correlators are familiar objects. Scattering processes in Lorentzian AdS/CFT that have normalizable external modes are known as transition amplitudes \cite{BalasubramanianKLT98,BalasubramanianKL98,BalasubramanianGL99,Raju10,Raju12A,Raju12B}, and are a special case of on shell correlators. In flat space, on shell correlators in momentum space are proportional to the S-matrix.

To implement a curved-space version of momentum conservation, we introduce enumerated differential operators $d_i$ defined as $d_i^\mu G^{\pm}(x_j,x) \equiv \delta_{ij}\nabla_x^\mu G^{\pm}(x_i,x) $. These operators are notational tools defined only at the integrand level. In flat space we have $d_i^\mu = -\nabla_{x_i}^\mu$ by translation invariance, but the relation between $d_i^\mu$ and $\nabla_{x_i}^\mu$ in general backgrounds is less simple. We show that on shell correlators can be written in terms of $d_i$ by using a contact diagram representation. For scalars, the on shell correlator is encoded by a function $A\left( (d_i)^n,(d_i)^{-m} \right)$ according to
\begin{equation}
\braket{\phi(x_1) \ldots \phi(x_k)}_{\text{on shell}} = \int d^{d+1}x \sqrt{-g} A\left( (d_i)^n,(d_i)^{-m} \right)  \prod_i^k G^{\pm}(x_i,x).
\end{equation}
Trivially, flat and curved space differ at the integrand level by commutators of $\nabla$'s and $\square^{-1}$'s. The contact representation simply allows us to use this fact to perform computations that parallel those for the S-matrix.\footnote{In spacetimes with symmetry, other such representations exist. Mellin space and embedding space are well-known examples.} For example, $\sum_i d_i = 0$ in the contact representation up to commutators $[d_i^\mu,d_i^\nu] \sim [\nabla^\mu,\nabla^\nu]$ that can be computed. Again up to commutators, the on-shell condition $(d_i)^2 = 0 $ holds for massless external scalars because the external legs are on shell. In simple cases, we find that the commutators can be interpreted as curvature couplings in the Lagrangian. Using the contact representation, it is clear that the on shell correlator recovers the S-matrix automatically in the flat-space limit. While we focus on four points, we give a few higher-point examples of on shell correlators in the contact representation. In short, the contact representation is a background-independent tool for studying weakly coupled QFT.

We use the contact representation to study the nonlinear sigma model (NLSM), which in flat space obeys flavor-kinematics duality and double copies to the special Galileon (SG). Promoting the standard formulation of the NLSM\footnote{For other representations, see \cite{CheungS16, CheungM21}.} to curved space, we find that color-kinematics duality holds for the NLSM at four points in generic spacetimes. The four-point BCJ relation is satisfied as well. Double copy operations encounter ambiguities, but we find a double copy procedure that connects the biadjoint scalar theory and the NLSM in general backgrounds. A similar procedure applied to AdS transition amplitudes relates the NLSM to a known SG result at four points. Everything that we discuss for on shell correlators applies to time-ordered correlators as well, albeit up to pinch terms. 

An outline of this note is the following. In Section 2, we review color-kinematics duality in flat space. In Section 3, we introduce the ingredients needed to implement color-kinematics duality in the position space integrand. In Section 4, we explore color-kinematics duality and a double copy procedure. We suggest future directions in Section 5.

We will use the notation and conventions from \cite{MeltzerS20} and work in mostly plus signature. Overall normalizations can be fixed by applying our approach in flat space, where comparison to momentum-space expressions is straightforward.

\section{Flat space warm-up: S-matrix review}

In this section, we review necessary details of the double copy for the NLSM. Unless stated otherwise, we will assume flat-space on-shell massless kinematics, $p_i^2 = 0$ and $\sum_i p_i = 0$, which we refer to as S-matrix kinematics for short. In preparation for curved space, we will identify statements that are valid more generally where appropriate.

\subsection{Color structure}

We review the color structure of NLSM amplitudes. The four-point amplitude $\mathcal{A}$ can be decomposed into color-ordered amplitudes as \footnote{For review, see for example \cite{Cheung17, Dixon13,ElvangH13}.}
\begin{equation}
\mathcal{A}(1^{a_1} 2^{a_2} 3^{a_3} 4^{a_4}) = \sum_{\sigma \in \mathcal{P}(2,3,4)} \text{Tr}\left( T^{\sigma(a_1)}T^{\sigma(a_2)}T^{\sigma(a_3)} T^{\sigma(a_4)} \right) A(\sigma(1) \sigma(2) \sigma(3) \sigma(4)),
\end{equation} 
where $\mathcal{P}(2,3,4)$ is the set of permutations with the location of 1 held fixed and $a_i$ are color indices.\footnote{For simplicity, we misuse terminology slightly and refer to flavor as color even when discussing the NLSM.}.

We can also decompose $\mathcal{A}$ into color factors $c_i$, which are products of $SU(N)$ structure constants $f^{abc}$. With Mandelstams $s,t,u$,
\begin{equation}
\mathcal{A}(1^{a_1} 2^{a_2} 3^{a_3} 4^{a_4}) = \frac{c_s n_s}{s} + \frac{c_t n_t}{t} + \frac{c_u n_u}{u},
\end{equation}
where kinematic numerators $n_i$ are functions of momenta. Using the basis
\begin{equation}
c_s = f^{a_1 a_2 b} f^{b a_3 a_4},
~~~~~
c_t = f^{ a_1 a_4b }f^{b  a_3 a_2 },
~~~~~
c_u = f^{a_1 a_3 b} f^{b a_2 a_4},
\end{equation}
the color factors obey the Jacobi relation $-c_s+c_t+c_u=0$. 

Decomposing color factors into traces of $T^a$, for example $c_s = -\text{tr} \left( [T^{a_1},T^{a_2}][T^{a_3},T^{a_4}]\right)$, we can write color-ordered amplitudes in terms of numerators,
\begin{equation}
A(1234) = \frac{n_s}{s}+\frac{n_t}{t},
~~~~~
A(1342) = -\frac{n_s}{s}-\frac{n_u}{u},
~~~~~
A(1423) = -\frac{n_t}{t} +\frac{n_u}{u}.
\label{AmpsInTermsOfNums}
\end{equation}
Color-ordered amplitudes obey Kleiss-Kuijf (KK) relations \cite{KK}, which are a direct consequence of the color decomposition. The four point KK relation is $A(1234)+A(1342)+A(1423) = 0
$. Equivalently, the propagator matrix has rank two. 

What we have reviewed so far follows entirely from color structure. Similar statements apply to $n$-point tree amplitudes regardless of the representation of the amplitude, how we split terms between numerator and denominator, and whether we work in S-matrix kinematics. For example, the KK relations are automatically obeyed by correlation functions in curved space.

\subsection{Color-kinematics duality and the double copy}

Color-kinematics duality \cite{BCJ} is the statement that numerators can be chosen to obey the same Jacobi relations as the color factors, $-n_s+n_t+n_u = 0 $. Numerators are not unique, as the shift
\begin{equation}
n_s \rightarrow n_s - s \alpha (s,t,u), ~~~ n_t \rightarrow n_t +t \alpha (s,t,u), ~~~ n_u \rightarrow n_u - u \alpha (s,t,u),
\end{equation}
leaves $\mathcal{A}$ invariant, which follows from $-c_s+c_t+c_u = 0$. This freedom to choose numerators is known as generalized gauge freedom. The numerator shift changes the Jacobi relation as
\begin{equation}
-n_s+n_t+n_u \rightarrow -n_s +n_t + n_u + (s+t+u)\alpha(s,t,u).
\end{equation}
In S-matrix kinematics, the sum is invariant. If color-kinematics duality is not already satisfied, a generalized gauge transformation cannot enforce color-kinematics duality. In more general kinematics, $\alpha = -\frac{n_t+n_u-n_s}{s+t+u}$ satisfies the Jacobi relations automatically \cite{Momeni:2020vvr,JohnsonJP20}. Equivalently, we can use the independent numerator in \eqref{AmpsInTermsOfNums} to enforce $-n_s+n_t+n_u = 0$. 

With S-matrix kinematics, $-n_s+n_t+n_u =0 $ implies the BCJ relation $tA(1234) = u A(1342)$. BCJ relations are a known prerequisite to the double copy. Recently, this has been studied in the case of massive double copy \cite{Momeni:2020vvr,JohnsonJP20}, but has been observed for massless kinematics as well \cite{HuangJL13,Allic14}.

Now we review the biadjoint scalar theory (BAS), nonlinear sigma model (NLSM), and special Galileon theory (SG). These theories of massless scalars are related by the double copy. The BAS action is
\begin{equation}
S_{BAS} = \int \frac{1}{2} \phi^{a \bar{a}} \square \phi^{a \bar{a}} + \frac{\lambda}{3} f^{a_1 a_2 a_3} f^{\bar{a}_1\bar{a}_2\bar{a}_3} \phi^{a_1 \bar{a}_1}\phi^{a_2 \bar{a}_2}\phi^{a_3 \bar{a}_3}.
\end{equation}
At four points,
\begin{equation}
\mathcal{A}_{BAS} = \frac{c_s \bar{c}_s}{s} +\frac{c_t \bar{c}_t}{t} +\frac{c_u \bar{c}_u}{u}.
\end{equation}
This amplitude double copies by replacing one of $c_i, \bar{c}_i$ with numerators $n_i$ that satisfy color-kinematics duality. In this way, the BAS furnishes the graph structure that defines the numerator decomposition.

Our main focus will be the NLSM. Up to four points, the NLSM action is \cite{KampfNT13, DuF16}
\begin{equation}
\mathcal{L}_{NLSM} 
= (\partial\phi)^2 + \frac{1}{12}f^{a_1 a_2 b} f^{b a_3 a_4} \partial_\mu \phi^{a_1} \phi^{a_2} \phi^{a_3} \partial^\mu \phi^{a_4}+ \mathcal{O}(\phi^6),
\end{equation}
where $\phi$ transforms in the adjoint representation of a global flavor symmetry. 

Without using S-matrix kinematics, the four-point amplitude is
\begin{equation}
\mathcal{A}(1234) =
c_s 
( p_{14}  - p_{2 4} - p_{1 3} +p_{2 3} ) 
+ 
c_t (p_{1 2}  - p_{24} - p_{1 3} +p_{34} )
+ 
c_u (p_{1 4}  - p_{3 4} - p_{1 2} +p_{23}).
\label{NLSMFlatAmplitude}
\end{equation}
where $p_{ij} \equiv  p_i\cdot p_j$, and the partial amplitudes are
\begin{align*}
A(1234) &=  -p_{14} +  2p_{2 4} + 2 p_{1 3} - p_{2 3} -p_{1 2} - p_{4 3},
\\
A(1342) &= 2 p_{14}  - p_{2 4} - p_{1 3} +2 p_{2 3}  - p_{3 4} - p_{1 2},
\\
A(1423) &= -p_{1 4}  + 2 p_{3 4} + 2 p_{1 2} -p_{3 2}- p_{4 2} - p_{1 3},
\numberthis
\end{align*}
which are proportional to $u,t,s$ respectively using S-matrix kinematics. The KK relation is satisfied by construction. These expressions will be useful in curved space.

Returning to S-matrix kinematics, the kinematic Jacobi relation is satisfied,
\begin{equation}
-n_s+n_t+n_u = 0.
\end{equation}
The double copy of the NLSM is the SG. The lowest-point vertex in the SG is $(\partial \phi)^2 (\partial \partial \phi)^2$ on shell ($\square \phi = 0$). The four-point SG amplitude is proportional to $stu$ in S-matrix kinematics and in this case, the double copy $c_i \rightarrow n_i$ of the NLSM amplitude gives
\begin{equation}
\mathcal{A}_{NLSM^2} = \mathcal{A}_{SG}.
\end{equation}

\section{Curved space: position-space integrand tools}

We now present the tools we use to explore color-kinematics duality in the position space integrand. We introduce differential operators, a contact representation, and correlators with external legs placed on shell. This approach allows us to compare S-matrix features to curved space quantities that are on a similar footing.  We will often use the compact notation $\int_{\mathcal{M}} d x  \equiv \int \sqrt{-g(x)} ~ d^{d+1} x$, and $\delta(x_1,x_2) \equiv \delta^{(d+1)}(x_1,x_2)$. The background metric is left unspecified.

\subsection{Wightman vs. Feynman for two-point functions}

The differences between the time-ordered and Wightman functions will be important, and we review those here. Taking $t_1 > t_2$, the Wightman and time-ordered functions are
\begin{equation}
 G^{+}(x_1,x_2)  = \braket{\phi(x_1) \phi(x_2)},
~~~~~~
 G^{-}(x_1,x_2)  = \braket{\phi(x_2) \phi(x_1)},
~~~~~~ 
G_F(x_1,x_2)  = \braket{T( \phi(x_1) \phi(x_2))},
\end{equation}
where $T$ denotes time-ordering. These are related by
\begin{equation}
G_F(x_1,x_2) = \theta(t_1-t_2)\braket{\phi(x_1) \phi(x_2)} +\theta(t_2-t_1)\braket{\phi(x_2) \phi(x_1)}.
\end{equation}
As we work in perturbation theory, we will need the time ordered, or Feynman, propagator and Wightman two-point function. In flat space, these are
\begin{align}
 G_F(x_1,x_2) & = \int d^{d+1}p ~ \frac{-i e^{i p\cdot (x_1-x_2)}}{p^2+m^2-i\epsilon},
\\
 G^{\pm}(x_1,x_2) & = \int d^{d+1}p ~ 2\pi \delta(p^2+m^2)\theta(\pm p^0)e^{i p \cdot (x_1-x_2)}.
\end{align}
Wightman functions appear in the context of the S-matrix. A unitarity cut amounts to the replacement $G_F(p) \rightarrow G^+(p)$ for cut lines. A similar statement holds for correlators in curved space, where the same replacement computes the unitarity cuts that enter in the curved-space optical theorem\footnote{As summarized in \cite{MeltzerS20}, the so-called CFT optical theorem in fact applies to causal, unitary QFTs on backgrounds in which energy is conserved. This optical theorem is a nonperturbative statement. The connection to cut diagrams, on the other hand, holds perturbatively and in \cite{MeltzerS20} was explored in AdS/CFT.} \cite{MeltzerS20}. In S-matrix language, the unitarity cut is an internal cut and LSZ cuts all external lines. While cuts are often implemented in momentum space, the cut also amounts to the replacement $G_F \rightarrow G^{+}$ in position space, which enters for example into the derivation of the largest-time equation in a general background (see \cite{MeltzerS20} for review).

Finally, the free two-point functions obey
\begin{equation}
\left( \square_{x_1} + m^2 + \xi R\right) G^{\pm}(x_1,x_2) = 0,
\label{EOMWightman}
\end{equation}
and 
\begin{equation}
\left( \square_{x_1} + m^2 + \xi R\right) G_F(x_1,x_2) = -i (-g(x_1))^{-\frac{1}{2}}\delta (x_1,x_2) 
\label{EOMFeynman}
\end{equation}
where $R$ is the Ricci scalar and $\xi$ is a free parameter. In this note, we study minimally-coupled scalars ($m = \xi = 0$) unless otherwise specified.

\subsection{Time-ordered correlators}
We begin by studying time-ordered correlation functions. Consider a diagram that appears in the NLSM.
\begin{equation}
\braket{T(\phi(x_1) \phi(x_2) \phi(x_3) \phi(x_4))}  = \int_{\mathcal{M}} d x   \nabla^\mu G_F(x_1,x) \nabla_\mu G_F(x_2,x)G_F(x_3,x)G_F(x_4,x) +\text{(perms)},
\label{aae}
\end{equation}  
where (perms) are the other diagrams that contribute, which we will leave implicit for the remainder of this discussion.

In flat space, momentum conservation at each vertex follows from translation invariance. In position space, a similar statement holds under integration by parts (IBP) provided the boundary terms vanish. 
\begin{align*}
\braket{T(\phi(x_1) \phi(x_2) \phi(x_3) \phi(x_4))} 
=-\int_{\mathcal{M}} d x  
&\left(
\square G_F(x_1,x)G_F(x_2,x)
+\nabla^\mu G_F(x_1,x) \nabla_\mu \right)
\\
&~~~~~~~
G_F(x_3,x)G_F(x_4,x).
\numberthis
\label{aag}
\end{align*}
When working in curved space, we will assume cluster decomposition so that the boundary terms coming from IBP are zero. Using \eqref{EOMFeynman},
\begin{align*}
\braket{T(\phi(x_1) \phi(x_2) \phi(x_3) \phi(x_4))} 
=&i G_F(x_2,x_1)G_F(x_3,x_1) G_F(x_4,x_1)
\\
&~~
-\int_{\mathcal{M}} d x  
\nabla^\mu G_F(x_1,x) \nabla_\mu 
 \left(G_F(x_3,x)G_F(x_4,x) \right) 
.
\numberthis
\label{aah}
\end{align*}
The first term arises because external lines are off shell, that is, are Feynman propagators instead of Wightman functions.\footnote{In flat space, the previous steps are $p_1\cdot p_2 = -p_1^2-p_1\cdot(p_3+p_4)$ in momentum space.} The term $G_F(x_2, x_1)G_F(x_3, x_1)G_F(x_4,x_1)$ is not singular as $x_{ij}^2 \rightarrow 0$ except for $x_{i1}^2 \rightarrow 0$. In this sense, these are partially disconnected terms, and are commonly known as pinch terms. Aside from such terms, IBP gives the same relations between derivatives that follows from momentum conservation for the S-matrix.

In flat space, introducing terms like $s/s$ allows us to decompose contact diagrams into a numerator representation. We can do the same in curved space by using \eqref{EOMFeynman}. Explicitly, we can write a contact diagram appearing in a $ \phi^4 $ theory as
\begin{equation}
\braket{T(\phi(x_1) \phi(x_2) \phi(x_3) \phi(x_4))} =i\int_{\mathcal{M}} d x ~d y  ~G_F(x_1,x)  G_F(x_2,x)
\square_{x} G_F(x,y)
G_F(x_3,y)G_F(x_4,y),
\label{aada}
\end{equation}
where $\square_x G_F(x,y) = \square_y G_F(x,y)$.

We can also use IBP to move $\square_x$ onto the external propagators\footnote{$(\square A) B = A (\square B)$ under IBP.}. This gives $\square_i G_F(x_i,x)$ terms, which are partially-disconnected contributions. Note that the partially disconnected terms here still contain an integral, but nevertheless have the same structure we identified earlier: each term lacks certain coincident-point singularities. For example, the sum
\begin{align*}
\braket{T(\phi(x_1) \phi(x_2) \phi(x_3) \phi(x_4))}_{\text{part. disc.}}
=&
i G_F(x_1,x_2)\int_{\mathcal{M}} d y ~G_F(x_1,y)G_F(x_3,y)G_F(x_4,y)
\\
&
i G_F(x_2,x_1)\int_{\mathcal{M}} d y ~ G_F(x_2,y)G_F(x_3,y)G_F(x_4,y)
\numberthis
\label{aah2}
\end{align*}
has  $x_{1 i}^2 \rightarrow 0, x_{2i}^2 \rightarrow 0$ singularities, but neither term does individually. These are proportional to $p_i^2 \neq 0$ in momentum space, and are zero in the S-matrix.

\subsection{On shell correlators}

The time-ordered correlator is an off-shell quantity, but there exists a closely-related object in which external fields have been placed on shell in a certain sense. We call this object the on shell correlator, $\braket{\phi(x_1) \phi(x_2) \phi(x_3) \phi(x_4)}_{W}$. For our purposes, it will be sufficient to define on shell correlators only at tree level in a weakly coupled QFT. We define on shell correlators as tree-level time-ordered correlators in which each external leg $G_F(x_i,x)$ has been replaced by a Wightman function $G^{\pm}(x_i,x)$,
\begin{equation}
\braket{\phi(x_1) \ldots \phi(x_n)}_{W} \equiv \braket{0| T(\phi(x_1) \ldots \phi(x_n))|0} \bigg |_{G_F(x_i, x) \rightarrow G^{\pm}(x_i,x)},
\end{equation}
where $x$ stands for locations of internal vertices. On shell correlators can also be obtained by integrating the amputated time-ordered correlator against $G^{\pm}$ in analogy with LSZ, 
\begin{equation}
\braket{\phi(x_1) \ldots \phi(x_n)}_{W} = \int  \left( \prod^n_i  dx_i' G^{\pm}(x_i,x_i') \right) \braket{0| T(\phi(x_1') \ldots \phi(x_n'))|0}_{amputated},
\end{equation}
or equivalently using the identity 
\begin{equation}
G^{\pm}(x_i,x) = i \int_{\mathcal{M}} dx' G^{\pm}(x_i,x_i') \square_{x_i'} G_F(x_i',x)
\end{equation}
on each external leg.

On shell correlators must contain at least one $G^{+}$ and one $G^{-}$, or energy conservation implies that the result vanishes. We will leave the specific choice implicit by using the notation $G^{\pm}(x_i,x)$. In the language of cuts, on shell correlators are Feynman diagrams in which each external line has been cut \cite{MeltzerS20}. While on shell correlators obey the same equations of motion as Wightman functions at tree level, it is unclear whether a general relation exists, although see \cite{Ostendorf84,Steinmann93,Steinmann95}.

While on shell correlators have not been studied in the literature in generality, they are familiar quantities in the specific cases: transition amplitudes are on shell correlators. Recall that in LSZ reduction, the S-matrix is proportional to the leading large-time singularity of the Fourier transform of $\braket{T(\phi(x_1)\ldots \phi(x_n))}$. This singularity corresponds to sending all external legs $G_F \rightarrow G^{\pm}$, which is the on shell correlator in momentum space\footnote{By the same argument, the Fourier transform of the on shell correlator is singular, and must be defined with care. It is natural to study the coefficient of this singularity, which is the S-matrix. We work in position space throughout and so do not encounter or address this issue.}. Concretely, suppose we have $n$ legs with $G^+(x_i,x)$ and $m$ with $G^-(x_i,x)$. The corresponding on shell correlator is related to the S-matrix $S$ and the momentum-space singularity of the time ordered correlator as
\begin{align*}
\prod_{i}^n \prod_j^m &\int dx_i e^{i p_i x_i} \int dy_j e^{-i q_j y_j}
\braket{T\left(\phi  (x_1) \ldots \phi (x_n) \phi (y_1) \ldots \phi (y_m) \right)}
\\
&\sim \prod_{i}^n \prod_j^m \frac{1}{p_i^2+m_i^2 -i\epsilon} \frac{1}{q_j^2+m_j^2 -i\epsilon}
\Bigg|_{\substack{p_i^2+m_i^2 \rightarrow 0\\ q_j^2+m_j^2 \rightarrow 0}}
\braket{p_1 \ldots p_n|S|q_1\ldots q_m}
\\
&
=
\braket{\phi^- (p_1) \ldots \phi^-(p_n) \phi^+(q_1) \ldots \phi^+(q_m)}_{W}.
\numberthis
\end{align*}
The $\phi^{\pm}$ are contracted with external legs $G^{\pm}$ respectively.

On shell correlators have been studied in curved space as well, specifically in AdS/CFT. Taking external points $x_i$ of bulk correlators to the boundary defines transition amplitudes in the Poincare patch, and are well-defined in the CFT \cite{BalasubramanianKLT98,BalasubramanianKL98,BalasubramanianGL99,Raju10,Raju12A,Raju12B}. On shell correlators in generic spacetime arise implicitly in the context of curved space unitarity methods \cite{MeltzerS20}.

On shell correlators generalize the on-shell property of S-matrix kinematics for the external lines. Specifically, these correlators do not contain the partially-disconnected terms that we found in the correlator. For instance the analogue of \eqref{aah} is
\begin{equation}
\braket{\phi(x_1) \phi(x_2) \phi(x_3) \phi(x_4)}_W 
=
-\int_{\mathcal{M}} d x  
\nabla^\mu G_F(x_1,x) \nabla_\mu 
 \left(G_F(x_3,x)G_F(x_4,x) \right),
\end{equation}
and similarly for \eqref{aah2}. We will study on shell correlators unless otherwise specified. Our statements will also apply to time-ordered correlation functions up to partially disconnected terms, such as those in \eqref{aah2}.

\subsection{$d_i$ in curved space}

In order to label external fields at the level of the position-space integrand, we introduce enumerated differential operators. For our purposes, these operators can be regarded as a notational device for covariant derivatives that keeps track of which function they act on. The $d_i$ play a similar role to the differential operators in \cite{EberhardtKM20, CasaliP20, DiwakarHRT21}. We define operators $d_i$ by their action on a scalar two-point function as
\begin{equation}
d_i^\mu   G(x_j,x) \equiv \nabla_x^\mu G(x_j,x) \delta_{i j},
\label{didefinition}
\end{equation}
where $G$ represents an external leg, either $G_F$ or  $G^{\pm}$. We define $d_{ij} \equiv d_i \cdot d_j$ for convenience. Derivatives acting on different legs commute, but can fail to commute when acting on the same leg,
\begin{equation}
[d_i^\mu,d_j^\nu] = [\nabla^\mu, \nabla^\nu] \delta_{ij}.
\end{equation} 
Explicitly, $[d_i^\mu,d_i^\nu]G(x_i,x) = 0$ but $[d_i^\mu,d_i^\nu]d_i^\rho G(x_i,x) \neq 0$. For spinning fields, we have $[d_i^\mu,d_i^\nu]G^{\rho\sigma}(x_i,x) \neq 0$ as well.

In flat space, $d_i = -\partial_{x_i}$ by translation invariance, and in momentum space this is $d_j = -i p_j$. We will not need an explicit representation of $d_i$ in curved space. However, note that $d_i$ depends on $x_i$ and $x$, and is a non-local differential operator in this sense.\footnote{It would be interesting to find an explicit representation for $d_i$, for instance in terms of the parallel propagator and $\nabla_{x_i}$.}

The background metric enters via commutators of $d_i$, for instance $[\nabla_\mu, \nabla_\nu]V^\rho = R^\rho_{~\lambda\mu \nu}V^\lambda$. In maximally-symmetric spacetimes, $R_{\rho \sigma \mu \nu} =\frac{R}{d(d+1)}\left(g_{\rho \mu}g_{\sigma \nu} - g_{\rho \nu}g_{\sigma \mu} \right)$ and $R_{\mu \nu } = \frac{R}{d+1}g_{\mu \nu}$, so the commutators become simple but remain non-zero. The following commutator will also be useful:
\begin{equation}
[\nabla_\rho, \square]  = ( [\nabla_\rho, \nabla_\mu ]\nabla^\mu+\nabla_\mu [\nabla_\rho,\nabla^\mu] ).
\end{equation}

\subsection{The $\square^{-1}$ operator}

In our approach, $\square^{-1}$ encodes the propagator. A formal expression for the propagator is
\begin{equation}
G_F(x,y) = -i \square_x^{-1} \left( \left(-g(x) \right)^{-\frac{1}{2}}\delta (x,y) \right).
\end{equation}
This property is satisfied by
\begin{equation}
\square_x^{-1}f(x) =  i\int_{\mathcal{M}} d  y f(y) G_F(x,y), 
\label{boxinverse}
\end{equation} 
which also obeys $\square_x \square^{-1}_x = \square^{-1}_x \square_x  = 1$. We assume that we can exchange the order of differentiation and integration. Enforcing uniqueness of $\square^{-1}$ as an operator requires specifying boundary conditions, while here we simply take \eqref{boxinverse} as the definition of $\square^{-1}$.

The definition \eqref{boxinverse} satisfies additional properties expected of $\square^{-1}$. Representing the integrand of an arbitrary tree diagram as $f_L(x) G_{F}(x,y) f_R(y) $,
\begin{align*}
\int_{\mathcal{M}} d x d y  f_L(x) G_{F}(x,y) f_R(y) 
&
=
-i \int_{\mathcal{M}} d x   d y   f_L(x) \square_x^{-1} \left(  \sqrt{-g(x)}^{-1}\delta(x,y) \right)f_R(y)
\numberthis
\\
&
=
-i \int_{\mathcal{M}} d x   d y   f_L(x) \square_y^{-1} \left( \sqrt{-g(y)}^{-1}\delta (x,y) \right)f_R(y),
\numberthis
\end{align*}
by definition. The expected IBP relation for $\square^{-1}$,
\begin{equation}
\int dx \left( A(x) (\square^{-1}_x B(x) ) - (\square^{-1}_x A(x)) B(x) \right) = 0,
\end{equation}
is also obeyed, 
\begin{align*}
\int_{\mathcal{M}} d x   d y   f_L(x) G_{F}(x,y) f_R(y) 
&=-i \int _{\mathcal{M}} d x    d y   f_L(x) \square_y^{-1} \left( \sqrt{-g(x)}^{-1}\delta (x,y) \right)  f_R(y) 
\\
&=-i \int_{\mathcal{M}} d x f_L(x) \square_x^{-1} f_R(x) = -i \int_{\mathcal{M}} d x \square_x^{-1}f_L(x)  f_R(x).
\numberthis
\label{InverseBoxIBP}
\end{align*}
Following \eqref{boxinverse}, the quantities in the second line are simply the LHS in different notation. We refer to the operation above as IBP for $\square^{-1}$. 

We can relate $\square^{-1}$ to $d_i$ as
\begin{equation}
\square^{-1}_x \prod_{k=i}^j G(x_k,x) = \square^{-1}_{i\ldots j} \prod_{k=i}^j G(x_k,x)
\end{equation}
because $\square^{-1}_{i \ldots j} (d_i +\ldots +d_j)^2 = 1$. For the same reason, this relationship holds when acting on external lines dressed with derivatives, for instance $d_i^\mu d_i^\nu \ldots d_i^\rho G(x_k,x)$. Similarly, we can also identify derivatives acting on $\square^{-1}$ with $d_i$ according to
\begin{equation}
\nabla_x^\mu \square^{-1}_{i \ldots j} = (d_i +\ldots+ d_j)^\mu \square^{-1}_{i \ldots j}.
\end{equation}
Note that $\nabla_x^\mu$ here does not act directly on external legs, but this identification is nevertheless consistent with $\nabla_x^2 \square^{-1}_{i \ldots j} = 1$.

While we lack translation invariance, it is at least formally possible to exchange $\nabla_x^\mu G_F(x,y)$ for $\nabla_y^\mu G_F(x,y)$ and account for the difference. First note that
\begin{equation}
[\nabla^\mu, \square^{-1}] = -\square^{-1}[\nabla^\mu,\square] \square^{-1}
\label{boxinversecommutatoridentity}
\end{equation}
 is an identity and so is automatically satisfied by \eqref{boxinverse}. Using \eqref{boxinversecommutatoridentity}, a derivative $\nabla$ acting at a vertex attached to only internal lines can be moved onto an external line and then written in terms of $d_i$. This implies $[\nabla_x^\mu,\square^{-1}_x] = [(d_i+\ldots+d_j)^\mu,\square_{i\ldots j}^{-1}]$ and gives the expected commutation rule for $d_i$.

The upshot of these properties of $d_i$ and $\square^{-1}$ is that we can carry out the following steps.
\begin{align*}
&\int_{\mathcal{M}} d x ~d y  G_F(x_1,x)  \nabla_x^\mu G_F(x_2,x)
 \nabla_{x,\mu} G_F(x,y)
G_F(x_3,y)G_F(x_4,y)
\\
&~~~~~~~~=\int_{\mathcal{M}} d x  G_F(x_1,x)  d_2^\mu G_F(x_2,x)
 (d_3+d_4)_\mu \square_{34}^{-1}
G_F(x_3,x)G_F(x_4,x),
\numberthis
\end{align*}
which is at least formally equal to
\begin{equation}
\int_{\mathcal{M}} d x G_F(x_1,x)  d_2^\mu G_F(x_2,x)
\left ([(d_3+d_4)_\mu, \square^{-1}_{34}] + \square^{-1}_{34} (d_3+d_4)_\mu\right)
G_F(x_3,x)G_F(x_4,x).
\label{commutatorwithboxinverse}
\end{equation}
In short, many of the same operations we can perform with $\nabla^\mu$ are also valid in terms of $d_i,d_i^{-1}$. The dependence on the curvature can be made explicit using \eqref{boxinversecommutatoridentity}.

What we have described so far also holds for a massive scalar as well. For non-minimal coupling, we use
\begin{equation}
\square_{i}^\xi  \equiv \square_{i} + \xi R,
\end{equation}
and the propagator $(\square_x^\xi)^{-1} $ defined as in \eqref{boxinverse} obeys the same properties as in the minimally-coupled case. 

Curvature terms $R$ that come from commutators of $d_i$ are associated with a single leg and so they commute with $d_j$ for $i\neq j$. To keep track of these in computations with $d_i$, we can promote $R \rightarrow R_i$ so that for example $d_i^\mu R_j = \delta_{ij} \nabla^\mu R $. While we mention this definition for completeness, $R$ will be constant in the examples we study.

\subsection{The contact representation}

The ingredients we have introduced imply that tree-level correlators admit an integrand representation that closely resembles the S-matrix. To summarize, an $n$-point tree diagram is entirely encoded by a function of $d_i,d_i^{-1}$ acting on an $n$-point zero-derivative contact diagram. This contact representation is defined as follows.
\begin{equation}
\braket{\phi(x_1) \ldots \phi(x_n)}
= 
\int_{\mathcal{M}} d x A(d_i, d_i^{-1})\prod_i^n G(x_i,x).
\end{equation}
The contact representation ``amplitude'' $A$ is a function of $d_i, d_{i}^{-1}$ and curvature terms in a specific ordering determined by the Feynman diagrams that enter. $G(x_i,x) = G_F(x_i,x)$ for time-ordered correlators and $G^{\pm}(x_i,x)$ for on shell correlators. 

In the contact representation, $d_i$ obey relations at each vertex that resemble momentum conservation in flat space. 
Concretely, in a diagram with $n$ legs, the outermost $d_i$ on any leg obeys 
\begin{equation}
\sum_i^n d_i = 0.
\end{equation}
This follows from IBP and the properties used in \eqref{commutatorwithboxinverse}. Specifically, the relation $\nabla_{x}^\mu \square^{-1}_{i \ldots j} =(d_i +\ldots+ d_j)^\mu \square^{-1}_{i \ldots j}$ allows us to convert $\nabla_x$ acting on propagators to the sum of $d_i$'s associated with the external legs on the other side of the propagator. The innermost $d_i$ acts on the external on-shell legs, and so obeys
\begin{equation}
(d_i)^2 = 0.
\end{equation}
In this way, the contact representation gives a partial generalization of S-matrix kinematics to curved space. The contact representation packages all differences between flat and curved space into commutators, which are proportional to background curvature. Once the on shell correlator is in the contact representation, it follows trivially that color-kinematics duality and any double copy procedure hold up to these curvature terms. It remains non-trivial whether these properties hold exactly.

We will often write the on shell correlator simply as $A$ and work at the integrand level. We spell out some simple examples to illustrate properties of $d_i$. A pure contact diagram is 
\begin{equation}
A_{\phi^n}(d_1,\ldots,d_n) = 1.
\end{equation}
The three-point function in a $(\partial \phi)^2 \phi$ theory is
\begin{equation}
A_{(\partial \phi)^2 \phi}(d_1,\ldots,d_3)
= 
d_{12} + d_{13}+d_{23}  = 0. 
\end{equation}
The four-point correlator for $\phi^3$ theory is
\begin{equation}
A_{\phi^3}(d_1,\ldots,d_4) = \frac{1}{\square_{12}}+\frac{1}{\square_{13}}+ \frac{1}{\square_{14}}.
\end{equation}
This representation is not unique, because for example $\square_{12}^{-1} =\square_{34}^{-1} $. A four-point derivative exchange correlator is
\begin{align*}
A_{(\partial\phi)^2 \phi}(d_1,\ldots,d_4) =& 
\frac{1}{\square_{12}}
(d_{12}-d_1\cdot (d_1+d_2) - d_2 \cdot (d_1+d_2) )
\\
&
(d_{34}-d_3\cdot (d_3+d_4) - d_4 \cdot (d_3+d_4) )
+\text{perms} 
\\
=& 
\frac{1}{\square_{12}} d_{12}d_{34} +\text{perms}  \propto d_{12}+d_{13}+d_{14}  = 0.
\numberthis
\end{align*}
Here, the IBP rule for $\square^{-1}$ specified that $\square^{-1}_{12}$ appeared on the left of $d_{12}$. As a reminder, we refer to this rule as IBP for $\square^{-1}$ but it is entirely equivalent to using \eqref{boxinverse} to simply replace the propagators with $\square^{-1}$.

To summarize, the contact representation at four points is found by using IBP to move all derivatives from propagators onto external legs, and then using IBP for $\square^{-1}$ to write the propagator as $\square^{-1}$ acting on external legs. The final result can be expressed in terms of $d_i, d_i^{-1}$. Higher point diagrams admit a contact representation as well. We give illustrative examples here. A zero derivative five-point diagram is
\begin{align*}
\braket{\phi(x_1) \ldots \phi(x_5)}
=&\int \left( \prod_i^3 dy_{i}\right )  G(x_1,y_1)G(x_2,y_1) G_F(y_1,y_2) G(x_3,y_2) G_F(y_2,y_3) 
G(x_4,y_3)G(x_5,y_3) ,
\\
A(d_1 \ldots d_1) =& 
\frac{1}{\square_{12}}  \frac{1}{ \square_{45}}.
\numberthis
\end{align*}
The same five point diagram with added derivatives is
\begin{align*}
\braket{\phi(x_1) \ldots \phi(x_5)}
=&\int \left( \prod_i^3 dy_{i}\right )  \left( G(x_1,y_1)G(x_2,y_1) \nabla_{y_1}^\mu G_F(y_1,y_2) \right)
\\
& (\nabla_{y_2,\mu}\nabla_{y_2,\nu} \nabla_{y_2,\rho} G(x_3,y_2)) \nabla_{y_3}^\nu \nabla_{y_3}^\rho G_F(y_2,y_3) 
G(x_4,y_3)G(x_5,y_3) ,
\\
A(d_1 \ldots d_1) =& 
(d_1 + d_2)^\mu \frac{1}{\square_{12}} d_{3,\mu} d_{3,\nu} d_{3,\rho}  (d_4 + d_5)^\nu (d_4 + d_5)^\rho \frac{1}{ \square_{45}}.
\numberthis
\end{align*}
At six points, we can have derivatives acting on vertices connected to only internal lines. First, consider the following six point diagram with no derivatives,
\begin{align*}
\braket{\phi(x_1) \ldots \phi(x_6)}
=&\int \left( \prod_i^4 dy_{i}\right )  \left( G(x_1,y_1)G(x_2,y_1) G_F(y_1,y_4) \right)
\\
&\left(G(x_3,y_2)G(x_4,y_2) G_F(y_2,y_4) \right)
\left(
G(x_5,y_3)G(x_6,y_3) G_F(y_3,y_4)  \right),
\\
A(d_1 \ldots d_6) =&
\frac{1}{\square_{12}} \frac{1}{ \square_{34}} \frac{1} {\square_{56}},
\numberthis
\end{align*}
where the ordering of $\square^{-1}_{ij}$ does not matter above because the operators commute. Derivatives on vertices $y_1 \ldots y_3$ can written in terms of $d_i$ using IBP. For instance, 
\begin{align*}
\braket{\phi(x_1) \ldots \phi(x_6)}
=&\int \left( \prod_i^4 dy_{i}\right )  \left( G(x_1,y_1) \nabla_{y_1}^\mu G(x_2,y_1) \nabla_{y_1,\mu}G_F(y_1,y_4) \right)
\\
&\left(G(x_3,y_2)G(x_4,y_2) G_F(y_2,y_4) \right)
\left(
G(x_5,y_3)G(x_6,y_3) G_F(y_3,y_4)  \right),
\\
A(d_1 \ldots d_6) &=  -\frac{1}{\square_{12}} (d_2^2 + d_1 \cdot d_2) \frac{1}{ \square_{34}} \frac{1} {\square_{56}}.
\numberthis
\end{align*}
The $\square_{34}^{-1}\square_{56}^{-1}$ still commute with all other terms. Now consider a diagram with derivatives on the vertex $y_4$,
\begin{align*}
\braket{\phi(x_1) \ldots \phi(x_6)}
&=\int \left( \prod_i^4 dy_{i}\right )   \left( G(x_1,y_1) G(x_2,y_1) \nabla_{y_4}^\mu G_F(y_1,y_4) \right)
\\
&\left(G(x_3,y_2)G(x_4,y_2) \nabla_{y_4,\mu} G_F(y_2,y_4) \right)
\left(
G(x_5,y_3)G(x_6,y_3) G_F(y_3,y_4)  \right),
\\
A(d_1 \ldots d_6) &=  (d_1+d_2)^\mu \frac{1}{\square_{12}} (d_3+d_4)_\mu \frac{1}{ \square_{34}} \frac{1} {\square_{56}}.
\numberthis
\end{align*}
Having shown examples of the contact representation above four points, we leave a systematic exploration of higher points to future work. 

\section{Color-kinematics duality at four points}

We now study color-kinematics duality in the contact representation. We work with theories that agree with the NLSM and the SG in the flat space limit.\footnote{For studies of the NLSM and the SG in curved space, see e.g. \cite{BonifacioHJR18,CheungMS20,FarnsworthHH21,DiwakarHRT21}. We thank K. Hinterbichler for comments on this.}. Through four points, we use the curved space theories
\begin{equation}
\mathcal{L}_{NLSM} 
= (\nabla\phi)^2 + \frac{1}{12}f^{a_1 a_2 b} f^{b a_3 a_4} \nabla_\mu \phi^{a_1} \phi^{a_2} \phi^{a_3} \nabla^\mu \phi^{a_4} ,
\end{equation}
and
\begin{equation}
\mathcal{L}_{SG} = (\nabla \phi)^2 +\frac{1}{12} (\nabla \phi)^2 (\nabla \nabla \phi)^2.
\end{equation}
As we study on shell correlators, off-shell terms like those containing $(\square \phi)^2$ will not appear at four points.

\subsection{Color-kinematics duality for NLSM}

We now implement color-kinematics duality for the NLSM in the contact representation. As the integrand is linear in $d_{ij}$, all $d_i$ obey S-matrix kinematics. This means we have many equivalent representations of the on shell correlator. As we shall see shortly, these different representations lead to inequivalent Jacobi relations and double copies. 

We begin with the representation that arises from direct application of the Feynman rules but without using further properties of $d_i$, as in \eqref{NLSMFlatAmplitude}.
\begin{equation}
A_{NLSM} =
c_s 
( d_{14}  - d_{2 4} - d_{1 3} +d_{2 3} ) 
+ 
c_t (d_{1 2}  - d_{4 2} - d_{1 3} +d_{4 3} )
+ 
c_u (d_{1 4}  - d_{3 4} - d_{1 2} +d_{3 2}).
\end{equation}
Introducing propagators $\square_{ij}^{-1}$ in different ways will lead to different numerators. It will instead be useful to generate numerators in a permutation invariant way, which is possible by introducing $\square^{-1}$ in the four-point interaction at the level of the action, performing Wick contractions, then converting to $d_i$. We will streamline this familiar process by defining the operator representation $\hat{A}$ of $A$ as follows. Using labels $a_i$ for color and $b_i$ for kinematics,
\begin{equation}
\int_{\mathcal{M}} dx \hat{A}_{NLSM} :\phi^{a_1,b_1} \phi^{a_2,b_2} \phi^{a_3,b_3} \phi^{a_4,b_4}:
\equiv  
\int_{\mathcal{M}} dx f^{a_1 a_2 e}f^{e a_3 a_4} d_{b_1b_4}:\phi^{a_1,b_1} \phi^{a_2,b_2} \phi^{a_3,b_3} \phi^{a_4,b_4}:.
\end{equation}
To obtain $A_{NLSM}$, we compute the correction to $\braket{T(\phi(x_1) \phi(x_2) \phi(x_3) \phi(x_4))}$ due to this interaction via Wick contraction and then replace the external legs with Wightman functions to obtain the on shell correlator. This procedure generates $A_{NLSM}$ without having to obtain it indirectly from the position-space integrand, and is entirely equivalent to working with the interaction vertex on shell. Labelling the external operators permits the use of $d_i$, and this procedure provides the Feynman rules for $d_i$.

Writing $\hat{A}_{NLSM}$ in numerator form,
\begin{equation}
\hat{A}_{NLSM}  
= 
f^{a_1 a_2 e}f^{e a_3 a_4} \frac{1}{\square_{b_1 b_2}} \square_{b_1 b_2}  d_{b_1 b_4}  .
\label{NLSMLagrangianNumeratorForm}
\end{equation}
The other three possible orderings are $\square_{b_1 b_2}   \square_{b_1 b_2}^{-1} d_{b_1 b_4},$ $ d_{b_1 b_4} \square_{b_1 b_2}   \square_{b_1 b_2}^{-1},  $ and $d_{b_1 b_4} \square_{b_1 b_2}^{-1} \square_{b_1 b_2}   $. We have checked that the ordering $d_{b_1 b_4} \square_{b_1 b_2}   \square_{b_1 b_2}^{-1}$ also gives color-kinematic dual numerators. The four-point case is somewhat ambiguous due to its simplicity, and so our aim is not to explore all possibilities, but only to demonstrate a working example. We will simply note that our choice is technically natural as it can be obtained by manipulations that are applicable to more general diagrams.

Now we take all possible contractions. The coefficient of $c_s$ is $(\square_{12}^{-1}\square_{12}+\square_{34}^{-1}\square_{34}) (d_{14}-d_{24}-d_{13}+d_{23})$, correctly reproducing $A_{NLSM}$. The presence of both $\square_{12}^{-1}$ and $\square_{34}^{-1}$ was inevitable, as neither is preferred. By comparison, the two are identical in flat space and so produce a single common denominator. We now have
\begin{align*}
A_{NLSM} =~~~& 
c_s \left(\frac{1}{\square_{12}}\square_{12}+\frac{1}{\square_{34}}\square_{34} \right)( d_{14}  - d_{2 4} - d_{1 3} +d_{2 3} ) 
\\
+& 
c_t \left(\frac{1}{\square_{14}}\square_{14}+\frac{1}{\square_{23}}\square_{23} \right) (d_{1 2}  - d_{24} - d_{1 3} +d_{34} )
\\
+ &
c_u\left(\frac{1}{\square_{13}}\square_{13}+\frac{1}{\square_{24}}\square_{24} \right)(d_{1 4}  - d_{3 4} - d_{1 2} +d_{23}).
\numberthis
\label{NLSMNumeratorForm}
\end{align*}
We identify numerators as the coefficient of $\square^{-1}$ in \eqref{NLSMLagrangianNumeratorForm}, which is equivalent to identifying each term to the right of a $\square^{-1}$ as a numerator even though $n_s, n_t, n_u$ do not have manifestly common denominators.
\begin{align*}
n_s &= (\square_{12}+\square_{34})( d_{14}  - d_{2 4} - d_{1 3} +d_{2 3} ) ,
\\
n_t &= 
(\square_{14}+\square_{23}) (d_{1 2}  - d_{24} - d_{1 3} +d_{34} ),
\\
n_u
&=
(\square_{13}+\square_{24})(d_{1 4}  - d_{3 4} - d_{1 2} +d_{23}).
\numberthis
\label{NLSMnums}
\end{align*}
Now we turn to the Jacobi relation. Because numerators are operators, we must specify the set of functions on which we evaluate $-n_s+n_t+n_u$. A natural choice is the contact representation. We implicitly make the same choice in the S-matrix case as well, which can be seen from the flat space limit of on shell correlators. In the contact representation, the Jacobi relation for numerators \eqref{NLSMnums} is satisfied,
\begin{equation}
-n_s+n_t+n_u 
=0.
\end{equation}
It turns out that $-n_s+n_t+n_u = 0 $ off shell as well.

The numerators have residual gauge freedom. For example, the shift
\begin{align*}
n_s &\rightarrow n_s - \alpha(d_i) d_{12},
\\
n_t &\rightarrow n_t +\alpha(d_i) d_{14},
\\
n_u &\rightarrow n_u +\alpha(d_i) d_{13},
\numberthis
\end{align*}
for any $\alpha(d_i)$ will leave the amplitude unchanged because $-c_s+c_t+c_u =0$. This numerator shift changes $-n_s+n_t+n_u$ by $\alpha(d_i)(d_{12}+d_{14}+ d_{13})$. For constant $\alpha$, the Jacobi relation is unchanged on shell. The on-shell numerators have an additional, off-shell freedom as well. Terms of the form $c_i f(d_1,\ldots,d_4) d_j^2$ are zero when added to the amplitude, and 
\begin{equation}
n_i \rightarrow n_i +\sum_j f_{ij}(d_1,\ldots, d_4) d_j^2
\end{equation}
leaves the numerators unchanged as well. In both cases, $d_j^2$ is the rightmost operator and so obeys $d_j^2 = 0$. However, these terms are important because they can affect the result of any double copy procedure in which a $d_j^2$ is no longer the rightmost operator after the entire numerator is squared.

The KK relations follow from color structure, and are automatically satisfied at $n$ points in any representation in curved space -- the contact representation and the fully integrated object included. BCJ relations however are not guaranteed. We find that the BCJ relations stated in \cite{BCJ} are satisfied when we choose the appropriate form of $A(ijkl)$. For example, using $A(1234) = d_{13}, A(1324) = d_{12}$, 
\begin{equation}
d_{12} A(1234) = d_{13} A(1324).
\end{equation}
As each term above contains at most two of the same $d_i$ and this acts on a scalar function, the $d_{ij}$ above all commute, and so the BCJ relations are satisfied.

\subsection{BAS double copy and squaring ambiguities}

It is not obvious whether differential operators admit a unique double copy procedure. The noncommutativity of $d_i$ and generalized gauge freedom of $n_i$ imply that we can obtain different final results from squaring the same set of $n_i$. Any differences are captured by commutators, which can be attributed to couplings to the curvature if the final result is permutation invariant. Carrying out a double copy procedure at the level of the action makes this equivalence explicit. In this section, we will give examples illustrating these points.

We first study a double copy from the BAS to the NLSM. The BAS four-point amplitude can be written as
\begin{equation}
A_{BAS} = c_s \bar{c}_s \left(\frac{1}{\square_{12}}+\frac{1}{\square_{34}} \right)
+
c_t \bar{c}_t \left(\frac{1}{\square_{14}}+\frac{1}{\square_{23}} \right)
+ 
c_u \bar{c}_u \left(\frac{1}{\square_{13}}+\frac{1}{\square_{24}} \right).
\label{BASForDC}
\end{equation}
We can replace $\bar{c}_s \rightarrow n_s$ in the above, but as $\bar{c}_s$ commutes with $d_i$, we must choose where to make this replacement. Different choices are inequivalent. If we make the replacement on the right, then we obtain the NLSM amplitude in \eqref{NLSMNumeratorForm}. This is equivalent to making the analogous replacement at the level of $\hat{A}_{BAS} = f^{a_1 a_2 b}f^{b a_3 a_4} f^{\bar{a}_1 \bar{a}_2 \bar{b}}f^{\bar{b} \bar{a}_3 \bar{a}_4} \square_{a_1 a_2}^{-1}$. In the end, this procedure gives a double copy.
\begin{equation}
A_{BAS \otimes NLSM}(d_1, \ldots, d_4) =A_{NLSM}(d_1, \ldots, d_4).
\end{equation}

Now we turn to double copy procedures for squaring the NLSM. We can carry out a double copy at the level of $\hat{A}_{NLSM}$ and the result will be permutation invariant. We begin with
\begin{equation}
\hat{A}_{NLSM}  
= 
f^{a_1 a_2 e}f^{e a_3 a_4} \frac{1}{\square_{b_1 b_2}} \square_{b_1 b_2}  d_{b_1 b_4}.\end{equation}
Two natural ways to square are 
\begin{equation}
\hat{A}_{(NLSM)^2_a}  
= 
\square_{b_1 b_2}  d_{b_1 b_4} \frac{1}{\square_{b_1 b_2}} \square_{b_1 b_2}  d_{b_1 b_4}
=
\square_{b_1 b_2}  d_{b_1 b_4}^2,
\end{equation}
and
\begin{equation}
\hat{A}_{(NLSM)^2_b}  
= 
\frac{1}{\square_{b_1 b_2}} \square_{b_1 b_2}  d_{b_1 b_4} \square_{b_1 b_2}  d_{b_1 b_4} 
=
d_{b_1 b_4} \square_{b_1 b_2}  d_{b_1 b_4}. 
\end{equation}
The two choices differ by $[\square_{b_1 b_2} ,d_{b_1 b_2}]d_{b_1 b_4}$. Trivially, any squaring procedure gives
\begin{equation}
\hat{A}_{(NLSM)^2_{a,b}} = \hat{A}_{SG} + (\text{curvature terms})_{a,b}.
\end{equation}
In practice, restoring the sum over contractions makes additional simplifications apparent and allows us to compute the curvature terms. Because this computation is done at the level of $\hat{A}_{SG}$, the curvature terms can be interpreted as additions to $\mathcal{L}_{NLSM}$. As their explicit form is not particularly illuminating and the action for SG in general spacetimes is not known for comparison, we omit them here.

\subsection{AdS transition amplitudes}

In this section, we give explicit expressions for a curved-space example, on shell correlators in AdS. We use notation from \cite{MeltzerS20}, to which we refer the reader for further details and overall numerical factors.

The metric of Poincare AdS is
\begin{equation}
ds^2 = \frac{dz^2 + \eta_{\mu \nu}dx^\mu dx^\nu}{z^2},
\end{equation}
where $z = 0$ is the AdS boundary. Bulk to bulk propagators in Poincare AdS$_{d+1}$ can written in $d$-dimensional momentum space in the flat (CFT) directions.
\begin{equation}
G_F(k,z_1,z_2) = (z_1 z_2)^h \int_0^\infty dp~p~ \frac{\mathcal{J}_\nu(p z_1) \mathcal{J}_\nu(p z_2)}{k^2+p^2-i\epsilon},
\end{equation}
where $\mathcal{J}_\nu(x)$ is the Bessel function of the first kind and $\nu = \Delta-h$. The bulk to bulk Wightman function admits a representation as a product of bulk-to-boundary Wightman functions \cite{MeltzerS20},
\begin{equation}
G^{\pm}(k,z_1,z_2) = (-k^2)^{-\nu}K^{\pm}(k,z_1)K^{\pm}(k,z_2),
\end{equation}
where $K$ is the bulk to boundary propagator and
\begin{equation}
K^{\pm}(k,z) = (-k^2)^{\frac{\nu}{2}}z^h \mathcal{J}_{\nu}(\sqrt{-k^2} z)\theta(k^2) \theta( \pm k^0),
\end{equation}
are the bulk to boundary Wightman functions. $G^\pm  \sim z^{\Delta}$ near the boundary, and so boundary terms arising from IBP generically vanish for on shell correlator. The bulk and boundary on shell correlators are
\begin{align*}
\braket{\phi(y_1)\phi(y_2)\phi(y_3)\phi(y_4)}_W
&=
\int \frac{d^d x dz }{z^{d+1}} A(d_i) \prod_j G^{\pm}(y_j,y) ,
\\
\braket{\mathcal{O}(x_1)\mathcal{O}(x_2)\mathcal{O}(x_3)\mathcal{O}(x_4)}_W
&=
\int \frac{d^d x dz }{z^{d+1}} A(d_i) \prod_j K^{\pm}(x_j,y) ,
\numberthis
\end{align*}
where bulk field $\phi$ is dual to $\mathcal{O}$ in the CFT and $y^\mu = (x,z)$. In both representations, color-kinematics duality for $A(d_i)$ follows automatically, as our earlier arguments were general.

Boundary on shell correlators in AdS/CFT are known as transition amplitudes \cite{BalasubramanianKLT98,BalasubramanianKL98,BalasubramanianGL99,Raju10,Raju12A,Raju12B}.\footnote{In the literature, transition amplitudes refer to diagrams with at least one leg on shell but other legs can be off shell. We focus on the case for which all legs are on shell.}. One way to define transition amplitudes between states in the CFT is to begin by tiling the cylinder with $n$ Poincare patches. Then, states $\bra{s}_n, \ket{s}_n$ of the $n$-th Poincare patch are defined by 
\begin{equation}
 \bra{s}_n \equiv { }_{n+1 }\bra{0} \mathcal{O}_{n+1}^-(p_1)\ldots \mathcal{O}_{n+1}^-(p_i) 
\end{equation}
and 
\begin{equation}
\ket{s}_n \equiv \mathcal{O}_{n-1}^+(q_1)\ldots \mathcal{O}_{n-1}^+(q_j) \ket{0}_{n-1 },
\end{equation}
which are the positive and negative energy modes on adjacent Poincare patches \cite{BalasubramanianGL99}. While our definition of on shell correlators was perturbative, this definition of transition amplitudes $\braket{p_1 \ldots p_i| q_1 \ldots q_j}$ is valid non-perturbatively. We refer the reader 
to \cite{BalasubramanianKLT98,BalasubramanianKL98,BalasubramanianGL99,Raju10,Raju12A,Raju12B} for a more detailed discussion.

We can compare this double copy to the Lagrangian derived in \cite{BonifacioHJR18} for the special galileon in $d+1=4$, in which the four-point on shell correlator comes from the interactions
\begin{equation}
\mathcal{L}_{SG } = 
-\frac{1}{2}(\nabla \phi)^2
-5 \phi^2
+\frac{127}{3}(\nabla \phi)^2 \phi^2
-\frac{23}{48}
(\nabla \phi)^4
-\frac{1}{24}
(\nabla \phi)^2 
 (\nabla \nabla \phi)^2.
\label{SG}
\end{equation}
Note that the Lagrangian above agrees on shell at four points with the Lagrangian in \cite{BonifacioHJR18}, but is otherwise different. We have simplified using the free equation of motion and set $l_{AdS} = 1$. Note that this form is not unique on shell. For example, a $R^3 \phi^4$ term can always be written as a $R^2 (\nabla \phi)^2 \phi^2$ term by using the equation of motion $\xi R\phi =  \square \phi$ and then integrating by parts. To check a double copy procedure, we need only compare to a Lagrangian that reproduces the correct SG four-point on shell correlator, rather than the off-shell Lagrangian that reproduces the correct time-ordered correlator.

\subsection{SG from a double copy}

In order to obtain \eqref{SG} via the double copy procedure we use, we need to implement color-kinematics duality with non-minimal coupling. We will do so in AdS$_d$. The Wightman functions now obey $\square^{\xi} G^{\pm} \equiv (\square - \xi R )G^{\pm} = 0$, or $d_i^2 G^{\pm}(x_i,x) = \xi R G^{\pm}(x_i,x)$. Repeating our earlier steps, we have numerators
\begin{align*}
n_s &= (\square_{12}^\xi+\square_{34}^\xi)( d_{14}  - d_{2 4} - d_{1 3} +d_{2 3} ) ,
\\
n_t &= (\square_{14}^\xi+\square_{23}^\xi) (d_{1 2}  - d_{24} - d_{1 3} +d_{34} ),
\\
n_u
&=
(\square_{13}^\xi+\square_{24}^\xi)(d_{1 4}  - d_{3 4} - d_{1 2} +d_{23}).
\numberthis
\end{align*}
This relation was satisfied off shell with $\xi =0$ and the terms proportional to $\xi R$ cancel algebraically. We therefore have
\begin{equation}
-n_s+n_t+n_u=0.
\end{equation}
Color-kinematics duality is therefore satisfied for any $\xi$.

We can use generalized gauge freedom to add certain terms with free coefficients to the color-kinematic numerators before squaring them. The modification
\begin{equation}
\hat{A}_{NLSM}=
f^{a_1 a_2 b}f^{b a_3 a_4} \frac{1}{\square_{b_1 b_2}^\xi} (\square_{b_1 b_2}^\xi d_{b_1 b_4}+ (c_1 d_{b_3 b_4}+c_2) (d_{b_4}^2+\xi R)),
\end{equation}
leaves the on shell correlator and numerators unchanged because $(d_{b_4}^2+\xi R) = 0$. Here $c_1$ is a numerical coefficient and $c_2 \sim 1/l_{AdS}^2$. As in the BAS example of double copy we gave earlier, we have a choice about how to include the second numerator. Following our procedure for the BAS case, we double copy by multiplication to the left, which gives
\begin{align*}
\hat{A}_{NLSM^2}&=\frac{1}{\square_{b_1 b_2}^\xi} (\square_{b_1 b_2}^\xi d_{b_1 b_4}+ (c_1 d_{b_3 b_4}+c_2) (d_{b_4}^2+\xi R))\square_{b_1 b_2}^\xi d_{b_1 b_4}
\\
&=
(d_{b_1 b_4} \square_{b_1 b_2}^\xi 
+ (c_1 d_{b_3 b_4}+c_2) (d_{b_4}^2+\xi R))d_{b_1 b_4}
.
\end{align*}
The first term gives the correct highest-derivative vertex $(\nabla \phi)^2 (\nabla \nabla \phi)$, which survives the flat space limit. We have checked that all possible lower-derivative vertices appear as well, and a basis of these interactions is $R^3 \phi^4, R^2(\nabla \phi)^2 \phi^2, R (\nabla \phi)^4$. As discussed earlier, $R^3 \phi^4$ can always be eliminated in favor of $R^2 (\nabla \phi)^2 \phi^2$ using the equation of motion, so reproducing \eqref{SG} requires matching the coefficients of $(\nabla\phi)^2 \phi^2$ and $(\nabla \phi)^4$ for some choice of $c_1,c_2$. The generalized gauge freedom shifts the $c_1=c_2=0$ result by
\begin{equation}
\Delta \hat{A}_{NLSM^2} =
(c_1 d_{b_3 b_4}+c_2 ) \frac{R}{d+1} d_{b_1 b_4}.
\end{equation}
This shifts the action by $c_1'  (\nabla \phi)^4 + c_2' (\nabla \phi)^2\phi^2 $ on shell where $c_1', c_2'$ are independent linear combinations of $c_1, c_2$. We therefore have sufficient freedom to reproduce the on shell four-point result derived from the $d=3$ Lagrangian \eqref{SG}, thereby implementing a double copy. 

Note that it was not obvious at the outset that there existed a shift of the numerators that enforced a double copy, left the amplitude invariant, and also left the Jacobi relations unchanged. While the numerators were not unique, only one set matched the SG result upon double copy. We picked a generalized gauge based on the desired double copy target, but it would be useful to identify a double copy procedure that gave the correct choice by construction.

\section{Discussion}
In this note, we developed tools to explore color-kinematics duality as a fundamental property of certain QFTs on any background. We took preliminary steps to apply these tools and implemented color-kinematics duality for the NLSM at four points. We also found a double copy procedure that connected the NLSM to the BAS theory in general spacetimes, and to the SG theory in AdS. While these examples were simple and contained some ambiguities, it is perhaps noteworthy that color-kinematics duality and double copy succeeded in any sense, given how general the setting was. We anticipate that ideas and ambiguities discussed here may be useful in seeding a more systematic investigation. We discuss some future directions.

One can apply our approach at higher points and then compare to the SG theory in AdS. It would also be interesting to explore KLT, because the singularity structure of correlators in curved space can be understood via cuts \cite{MeltzerS20}. A more detailed study of $d_i, d_i^{-1}$ and their commutators would be useful as well. We see no fundamental obstruction to continuing along these lines.

Our approach may be applicable to Yang-Mills theory and gravity, beginning even at three points. A natural proposal for the spinning contact representation of spin-1 fields, for example, involves Wightman functions $G^{\pm}_{\mu \nu}(x,y)$ and is 
\begin{equation}
\braket{\phi^{\mu_1} (x_1) \phi^{\mu_2}(x_2) \phi^{\mu_3}(x_3) \phi^{\mu_4}(x_4)}_{W}
= 
\int d^{d+1}x A^{\nu_1 \nu_2 \nu_3 \nu_4}(d_i)\prod_i^4 G^\pm_{\mu_i \nu_i}(x_i,x).
\end{equation}
For the contact diagram in Yang-Mills, $A^{\nu_1 \nu_2 \nu_3 \nu_4}$ is the quartic vertex obtained from the Feynman rules.

It would be interesting to compute scattering amplitudes perturbatively in the background metric in an asymptotically flat spacetime, where an S-matrix can still defined. This may help separate the dependence on background curvature from difficulties arising from not having a simple notion of on-shell external states. This may also determine whether the background must satisfy Einstein's equations, a question our approach is seemingly insensitive to.

The failure of derivatives to commute was a central feature of how background curvature was manifest in the contact representation. As this non-commutativity can be understood in the Lagrangian as well, it would be useful to understand if an unambiguous Lagrangian-level double copy procedure could fix these ambiguities. Lagrangian approaches to gravity that make twofold Lorentz symmetry manifest in curved spacetimes (e.g. \cite{CheungR16,CheungR17}) may probe such features.

Independently of our study of color-kinematics duality, we found that on shell correlators allowed us to study on-shell external configurations in settings without an S-matrix. On shell correlators themselves have received little attention in the literature, particularly outside of their connection to transition amplitudes in AdS. However, studying on shell correlators further may better enable extensions of flat-space methods to curved spacetimes, and in particular help us understand the origin of color-kinematics duality.

\section{Acknowledgements}
I am grateful to Clifford Cheung, David Meltzer, and Julio Parra-Martinez for many insightful discussions, and also for comments on the draft. I thank the Walter Burke Institute for Theoretical Physics for hospitality while this project was in its final stages. I have been supported by a National Science Foundation grant NSF-PHY/211673 and the College of Arts and Sciences of the University of Kentucky over the course of this work.

\bibliographystyle{ssg}
\bibliography{refs}

\end{document}